\documentclass[sn-mathphys]{sn-jnl}

\jyear{2021}%

\usepackage{eurosym}
\usepackage{graphicx}
\usepackage[color,matrix,arrow,all,2cell]{xy}
\usepackage[T1]{fontenc}

\newcommand*{\Scale}[2][4]{\scalebox{#1}{$#2$}}
\newcommand{\ba}{\begin{array}}
\newcommand{\ea}{\end{array}}
\newcommand{\up}{{\color{black}\rightarrow}}
\newcommand{\down}{{\color{black}\leftarrow}}

\renewcommand{\rho}{{\color{black}r}}
\renewcommand{\pi}{{\color{black}p}}

\begin{document}

\title{{\color{black}Multicyclic norias: a first-transition approach to extreme values of the currents}}

\author*[1]{\fnm{Matteo} \sur{ Polettini}}\email{matteo.polettini@uni.lu}

\author[2]{\fnm{Izaak} \sur{ Neri}}\email{izaak.neri@kcl.ac.uk}



\affil*[1]{\orgdiv{Department of Physics and Materials Science}, \orgname{ University of Luxembourg}, \orgaddress{\street{Campus Limpertsberg, 162a avenue de la Fa\"iencerie}, \city{Luxembourg}, \postcode{L-1511}, \country{G. D. Luxembourg}}}

\affil*[2]{\orgdiv{Department of Mathematics}, \orgname{ King’s College London}, \orgaddress{\street{Strand}, \city{London}, \postcode{WC2R 2LS}, \country{UK}}}

\abstract{{\color{black} 
For continuous-time Markov chains we prove that, depending on the notion of effective affinity $F$, the probability of an edge current to ever become negative is either $1$ if $F< 0$ else $\sim \exp - F$. The result generalizes a ``noria'' formula to multicyclic networks. We give operational insights on the effective affinity and compare several estimators, arguing that stopping problems may be more accurate in assessing the nonequilibrium nature of a system according to a local observer. Finally we elaborate on the similarity with the Boltzmann formula. The results are based on a constructive first-transition approach.}}  


\maketitle

\newpage

\section{Introduction}

{\color{black} 

Let us first provide an illustrative simple example of some of our main results. Perform a random walk on the network
$$\xymatrix{\star \ar@/^2pt/@{->}[rr]^{+\text{\euro}} \ar@/_2pt/@{<-}[rr]_{-\text{\euro}} & & \bullet \ar@{-}[dd] \\ \\ \bullet \ar@{-}[uu]   & & \bullet \ar@{-}[ll] \ar@{-}[uull]}$$
where every time you cross one specific edge in one direction you get 1\euro, and when you cross it in the opposite direction you pay 1\euro. All other transitions also give and take credit, but in liras. Initially your pockets are empty of euros, and while you have a virtually infinite reservoir of liras your bank would not convert them into euros.

The question we pose and answer in this paper is: assuming that you live forever, what is the probability $\mathfrak{f}_-$ that you eventually get broke (get to -1\euro, that you cannot actually afford to pay)?

In the special case where you are initially in $\star$ (so that at least you get a chance to not get immediately broke!), we find, on the assumption $F > 0$,
\begin{align}
\mathfrak{f}_- = \exp - F, \label{eq:norianoria}
\end{align}
with
\begin{align}
F = \log \frac{
\Scale[0.8]{
  \ba{c}\xymatrix{ & \ar[d] \ar@{<-}[l]  \\ \ar[u] & \ar[l]} \ea 
+ \ba{c}\xymatrix{ & \ar@{->}[d]   \ar@{<-}[l]\\ \ar[u] & \ar@{->}[ul] } \ea
+ \ba{c}\xymatrix{   & \ar[d]   \ar@{<-}[l] \\  \ar[r] & \ar@{->}[ul] } \ea } }{ 
\Scale[0.8]{
 \ba{c}\xymatrix{ & \ar@{<-}[d] \ar@{->}[l] \\ \ar@{<-}[u] & \ar@{<-}[l]} \ea
+ \ba{c}\xymatrix{ & \ar@{<-}[d] \ar@{->}[l]  \\ \ar[u] & \ar@{<-}[ul] } \ea
+ \ba{c}\xymatrix{   & \ar@{<-}[d] \ar@{->}[l]  \\  \ar[r] & \ar@{<-}[ul]  } \ea
 } },
\end{align}
where each diagram implies multiplication by the corresponding rates of the Markov chain. If instead $F < 0$, the probability to go bankrupt is $1$.

In the case where the diagonal transition has vanishing rates in both directions, in the last expression the diagrams containing diagonal terms disappear and we are left with the log-ratio of two cyclic contributions
\begin{align}
\log \frac{  \Scale[0.8]{ \ba{c}\xymatrix{ & \ar[d] \ar@{<-}[l]  \\ \ar[u] & \ar[l]} \ea  }}{\Scale[0.8]{ \ba{c}\xymatrix{ & \ar@{<-}[d] \ar@{->}[l] \\ \ar@{<-}[u] & \ar@{<-}[l]} \ea.
 }}  = A.
 \end{align}
This object is known as cycle affinity, a measure of the probability of performing the cycle in one direction relative to the opposite. For unicyclic networks (``norias'') Eq.\,(\ref{eq:norianoria}) boils down to a remarkable formula first obtained by Bauer and Cornu \cite{bauer2014affinity} (which in fact is slightly more general in that the initial state can be chosen arbitarily). As reasonable, the cycle is completed more often in the favourable direction $(A > 0$) rather than the unfavourable one; then the Bauer-Cornu formula yields the probability of the rare event of the cycle to ever be completed more often in the unfavourable direction. Generalizations have been proven for the total entropy production (a weighted balance of euros and liras) using the tools of martingale theory \cite{neri2017statistics,neri2019integral,neri2022universal}\footnote{More precisely, from Eq.\,(10) in Ref.\cite{neri2017statistics}, the stochastic entropy production of a generic system satisfies $\mathfrak{f}_- = 1/e$ ($e$ Neper's number), which is the above formula for $A = 1$; equivalently take $s_+ \to \infty$ and set $s_-=1$ in Eq.\,(5) in Ref.\,\cite{neri2019integral}.}, and discussed in the light of first-exit time problems \cite{ptaszynski2018first}.

Beyond the probabilistic interpretation, cycle affinities also afford two different thermodynamic interpretations, one global and one local, of which we give here an intuitive sketch (but see Sec.\,\ref{sec:effaff} for details), on the assumption that forward and backward rates of a transition $x \leftrightarrow y$ have ratio $\exp \delta q_{xy} / T_{xy}$ with $\delta q$ the heat (in our example: currency) exchange along that transition and $T$ the temperature (in our example this could be the interest rate) of that transition, i.e. a measure of how inconvenient it is to perform that transition (to borrow money). Then the global one is as Carnot's entropy production along a cyclic process \cite{polettini2012nonequilibrium}
\begin{align}
A = \oint \frac{\delta q}{T}
\end{align}
where $\oint$ is a shorthand for the sum over cyclic transitions. This takes into account all contributions, e.g. from liras and from euros, even if for the problem at hand liras to not play any role. Thus this latter interpretation has little operational value. The local one is
\begin{align}
A = \left(\frac{1}{T^\text{\euro}} - \frac{1}{T^\varnothing} \right) \delta \text{\euro}, \label{eq:euro}
\end{align}
in terms the specific transaction in euros, its temperature $T^\text{\euro}$, and the value of such temperature $T^\varnothing$ at which forward and backward euro transactions happen with the same frequency (that is: the rest of the world does not care much of whatever you do with your euros).

Going back to our main result Eq.\,(\ref{eq:norianoria}) (which, we remind, holds for arbitrary non-unicyclic networks), here the effective affinity does not have such a simple global interpretation, but still retains the local interpretation. What is lost with respect to norias is the independence from the initial state: while it does not matter where you are initially along a cycle to complete that cycle, it does matter where you are in a network to perform an arbitrary composition of cycles that touch the initial state. We generalize Eq.\,(\ref{eq:norianoria}) appropriately.

We then show by computational examples that first-passage and extreme value problems such as the one above might give a better estimate of the effective affinity than do fluctuation and fluctuation-dissipation relations at fixed stopping time. Finally, Eq.\,(\ref{eq:norianoria}) is reminiscent of Boltzmann's formula, but turned upside down. We linger on this analogy towards the conclusions.
}

\section{Framework}

\subsection{State-space processes}

Consider an irreducible continuous-time Markov chain on finite state space $\mathcal{X}$. We characterize it in terms of the probability $p^\mathcal{X}_t = \{p^{\mathcal{X}}_t(x), x \in \mathcal{X}\}$ of being at $x$ at time $t$, which satisfies the continuous-time master equation 
\begin{align}
\frac{d}{dt} p^{\mathcal{X}}_t(x) = \sum_{x' (\neq x) \in \mathcal{X}} \left[ \rho(x\vert x') \,p^{\mathcal{X}}_t(x') - \rho(x'\vert x) \,p^{\mathcal{X}}_t(x) \right] \label{eq:one}
\end{align}
starting from a given initial distribution $p^{\mathcal{X}}_0$, with non-negative rates $\rho(x\vert x')$ of jumping from $x'$ to $x$. We also define the continuous-time (adjoint) generator $R$ with matrix entries
\begin{align}
R_{x,x'} = \rho(x\vert x') - \delta_{x,x'} r(x')  \label{eq:generator}
\end{align}
{\color{black} where  $\delta$ is Kroenecker's delta and
\begin{align}
r(x) = \sum_{x' \in \mathcal{X}} \rho(x'\vert x)
\end{align}
is the exit rate out of a state}. The master equation reads in vector form $\frac{d}{dt} p^{\mathcal{X}}_t = R p^{\mathcal{X}}_t$ and its stationary distribution solves $R p^{\mathcal{X}}_\infty = 0$. From now on we do not specify the range of summation unless necessary.

We now focus on a pair of connected states, namely $x,x' = 1,2$ without loss of generality. We assume that edge $1 \leftrightarrow 2$ is not a bridge, that is, that its removal does not disconnect the system, and denote $\mathcal{X}_\varnothing$ (or simply $\varnothing$) a system where  edge $1 \leftrightarrow 2$ is removed. Transitions between these states are deemed to be visible to an external observer. Let $\ell \in \mathcal{L} = \{ \up = 2 \to 1, \down = 1 \to 2\}$ denote transitions between such states, to and from, and $\mathtt{t}(\cdot), \mathtt{s}(\cdot) \in \{1,2\}$ the source and target states of a transition, i.e. $\mathtt{t}(\up) =  \mathtt{s}(\down) = 1$ and $\mathtt{t}(\down) = \mathtt{s}(\up) = 2$.

Letting $n(\ell)$ be the number of times transition $\ell$ occurs along a realization of the process, we define the visible activity and the cumulated current respectively as
\begin{align}
n & = n(\up) + n(\down), \\
c & = n(\up) - n(\down). \label{eq:cur}
\end{align}
Notice that they typically grow linearly in time; thus we denote the mean stationary current (i.e. cumulated current per unit time) as
\begin{align}
\langle \dot{c} \rangle = \rho(1\vert 2) p^\mathcal{X}_\infty(2) - \rho(2\vert 1) p^\mathcal{X}_\infty(1).  
\end{align}

Our final goal is to compute the probability $\mathfrak{f}_\pm$ that the cumulated current $c$ takes value $\pm 1$ at least once as the process unfolds from time $t = 0$ to time $t \to +\infty$. 

\subsection{Transition-state processes}

Our strategy is to lift the description of the process from state space $\mathcal{X}$ to transition space $\mathcal{L}$, following the treatment of Ref.\,\cite{harunari2022beat}. {\color{black} The central objects to be calculated are the trans-transition probabilities $p(\ell\vert \ell')$ that the next observable transition is $\ell$ given that the previous was $\ell'$. An intuitive way to go about this would by brute-force coarse-graining of a stochastic trajectory $\boldsymbol{x},\boldsymbol{\tau} = x_0,\tau_0 \to x_1,\tau_1 \to \ldots \to x_k,\tau_k$ in state space, where $x_j$ are the visited states and $\tau_j$ are the permanence times. Then $p(\ell\vert \ell')$ can be computed by marginalization of the probability density function at fixed number of jumps $k$
\begin{align}
f_k(\boldsymbol{x},\boldsymbol{\tau}) = \prod_{i = 1}^{k-1} r(x_{i+1}\vert x_i) \, e^{- r(x_i) \tau_i}
\end{align}
by integrating away the intermediate times and summing over all trajectories between the target state of $\ell'$ and the source state of $\ell$ but that otherwise do not contain observable transitions, and multiplying by the rate of this latter transition. This direct procedure is illustrated in the Appendix of Ref.\,\cite{harunari2022beat}.}

{\color{black} A more elegant line is the following.} Notice that with probability one the {\color{black} visible} activity takes any positive integer value {\color{black} and at any given time $t$ does not depend on future information}. Thus the time when the activity reaches a certain value $n$ for the first time is a valid stopping time. Then by the strong Markov property \cite{ibe2013markov} the {\color{black} event} of being at $x$ after $n$ visible transitions is also a Markov process in state space.
 Let $p^\mathcal{L}_n(\ell)$ be the probability that the $n$-th visible transition is $\ell$. Notice that the probability that the next transition is $\ell$ given that the previous was $\ell'$ only depends on the target state of $\ell'$. Thus we conclude that $p^\mathcal{L}_n$ satisfies a discrete-time Markov chain in transition space
\begin{align}
p^\mathcal{L}_{n+1}(\ell) = \sum_{\ell' \in \mathcal{L}} \pi(\ell\vert \ell') \, p^\mathcal{L}_{n}(\ell')
\end{align}
evolving from some initial probability $p^\mathcal{L}_1(\ell)$ that the first transition is $\ell$. The $\pi(\ell\vert \ell')$ are the the so-called trans-transition probabilities; we arrange them in a trans-transition matrix $P$ with entries $P_{\ell,\ell'} = \pi(\ell\vert \ell')$. 

Both the initial transition probability and the trans-transition probabilities can be obtained from the initial state probability $p^\mathcal{X}_0$ and the transition rates $\rho(x\vert x')$ by solving first-transition time problems. In particular the probability that, starting from $x$, $\ell$ is the first visible transition and that it occurs in the time interval $[t,t+dt)$ is given by
\begin{align}
\rho(\mathtt{t}(\ell)\vert \mathtt{s}(\ell)) \left[ \exp tS\right]_{\mathtt{s}(\ell),x} dt\label{eq:firstt}
\end{align}
where $S$ is the matrix obtained from $R$ by setting to zero the off-diagonal entries corresponding to the visible transition, namely
\begin{equation} \label{eq:S}
\begin{aligned}
S_{x,x'} & = R_{x,x'}, \qquad \mathrm{for}\,(x,x') \neq (1,2), (2, 1),  \\
S_{1,2}  = S_{2,1} & = 0 .
\end{aligned}
\end{equation}

By integrating Eq.\,(\ref{eq:firstt}) from $t = 0$ to infinity and evaluating at $x = \mathtt{t}(\ell')$ we find, for all $\ell,\ell' \in \mathcal{L}$, the trans-transition probabilities
\begin{align}
\pi(\ell\vert \ell') =  - \rho(\mathtt{t}(\ell)\vert \mathtt{s}(\ell)) [S^{-1}]_{\mathtt{s}(\ell) \mathtt{t}(\ell')},\label{eq:trantrans}
\end{align}
and, for all $\ell \in \mathcal{L}$, the probability of the first transition
\begin{align}
p^\mathcal{L}_1(\ell) = - \rho(\mathtt{t}(\ell)\vert \mathtt{s}(\ell)) \sum_{x} [S^{-1}]_{\mathtt{s}(\ell),x} \, p^\mathcal{X}_0(x). \label{eq:init}
\end{align}
It was proven \cite[Appendix]{harunari2022beat} that trans-transition probabilities and the initial transition probability are positive and normalized, as they should be:
\begin{align}
1 = p^\mathcal{L}_1(\up) + p^\mathcal{L}_1(\down) = \pi(\up\vert \ell) + \pi(\down\vert  \ell ), \quad \mathrm{for}\;\ell \in \mathcal{L}. \label{eq:norm}
\end{align}

Explicitly, the trans-transition matrix is given by
\begin{align}
P = \frac{1}{\nu_{\up} + \nu_{\down} - 
\nu_0}
\left(\ba{cc} \nu_{\up} - 
\nu_\circ &  \nu_{\up} \\ \nu_{\down} & \nu_{\down} - 
\nu_\circ \ea\right) \label{eq:explicit}
\end{align}
where, letting $A_{\setminus (x_1, \ldots , x_n \vert  x'_1, \ldots , x'_n)}$ be a matrix from which rows $x_1, \ldots , x_n$ and columns $x'_1, \ldots , x'_n$ are removed, we have
\begin{align}
\nu_\circ & = \rho(1\vert 2) \rho(2\vert 1) \det R_{\setminus (1,2 \vert  2, 1)} \nonumber \\
\nu_{\up} & = \rho(1\vert 2) \det R_{\setminus (2\vert 1)}  \label{eq:ten} \\
\nu_{\down} & = \rho(2\vert 1) \det R_{\setminus (1\vert 2)}.  \nonumber
\end{align}
A proof of these expressions is given in Appendix \ref{app1}.

\section{Results}

\subsection{Statement and derivation of the main result}

We can now formulate our problem of calculating the probability that the cumulated current ever hits value $-1$ (case $+1$ for later) as
\begin{align}
\mathfrak{f}_- = \sum_{n = 1}^\infty \mathfrak{f}^{(n)}_- \label{eq:sum}
\end{align}
where $\mathfrak{f}^{(n)}_-$ is the probability that the cumulated current $c$ takes value $-1$ for the first time at the $n-$th visible transition. The first is just the probability that the transition occurs right-away:
\begin{align}
\mathfrak{f}^{(1)}_- & =  p^\mathcal{L}_1(\down).
\end{align}
Notice instead that the cumulated current cannot be $-1$ after two  visible transitions:
\begin{align}
\mathfrak{f}^{(2)}_- & = 0.
\end{align}
For the cumulated current to be $-1$ for the first time at the third visible transition, we need that the first visible transition is $\up$ and the second and third are $\down$, therefore: 
\begin{align}
\mathfrak{f}^{(3)}_-  & =  \pi(\down\vert \down) \pi(\down\vert \up) p^\mathcal{L}_1(\up) .
\end{align}
To go beyond, first notice that all probabilities at even $n$ vanish. At odd $n$, we need to count all different paths of $2n+1$ steps that perform a $\down$ transition leading from $c = 0$ to $c = -1$ for the first time as the last step, and multiply each path by the corresponding probability. Namely, we need to count all different sequences $(\ell_1,\ell_2, \ldots, \ell_{2n+1})$ such that:
\begin{itemize}
\item[1)] $\ell_{1} = \up$ and $\ell_{2n+1} = \down$;
\item[2)] at any intermediate step the number of $\down$ is never greater than the number of $\up$ and at $2n$ the number of $\down$ is exactly equal to the number of $\up$;
\item[3)] they have a given amount $k \leq n$ of $\down\vert \up$ trans-transitions (which also fixes the number of $\down\vert \down$, $\up\vert \up$, and $\up\vert \down$ trans-transitions). 
\end{itemize}

\begin{figure}
\begin{center}
\includegraphics[angle = -90,width=10cm]{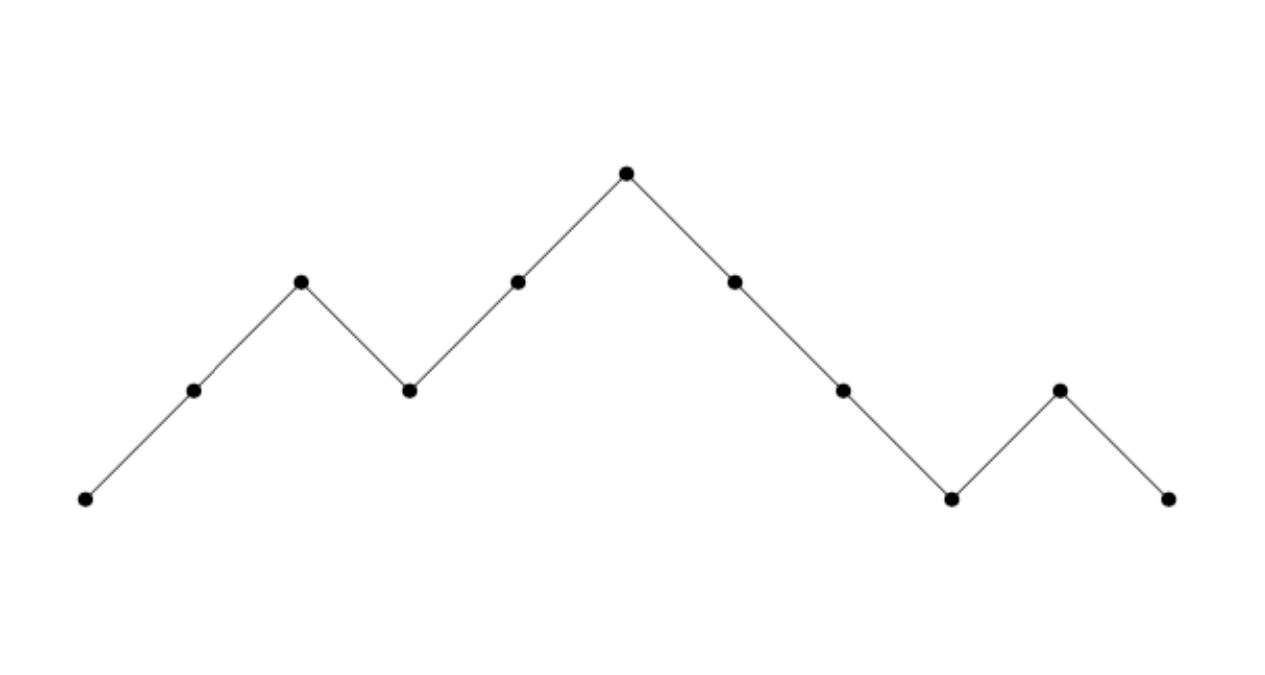}
\end{center}
\caption{A mountain with $n = 5$ up and down slopes, $3$ peaks and $2$ valleys.}\label{fig:mountain}
\end{figure}

In fact, this question maps to a well-known enumeration problem: if we replace $\up$ with a  $45^\circ$ unit segment and $\down$ with a $-45^\circ$ unit segment, we need to count all ``mountains'' of length $2n$ that can be drawn without lifting the pencil and that have exactly $k$ peaks (see Fig.\,1). This problem is well-known to be solved by the {\color{black} Narayana} numbers \cite{petersen2015eulerian}
\begin{align}
N(n,k) = \frac{1}{n} { n \choose k} { n \choose k-1}.
\end{align}
Therefore for $n \geq 1$ the result to our problem is
\begin{align}
\mathfrak{f}^{(2n+1)}_- = p^\mathcal{L}_1(\up) \frac{\pi(\down\vert \down)}{\pi(\up\vert \down)}  \sum_{k = 1}^n N(n,k) [\pi(\down\vert \down) \pi(\up\vert \up)]^{n-k} [\pi(\down\vert \up) \pi(\up\vert \down)]^{k}. \label{eq:crucial}
\end{align}
Notice that the prefactor in Eq.\,(\ref{eq:crucial}) accounts for the initial transition (which must be $\down$), for the last transition (which must be $\down$ given that the previous was also $\down$), and for the fact that in such mountains valleys are one less than peaks.

We now use the fact that {\color{black} Narayana} numbers admit the generating function \cite{stanley2011enumerative}
\begin{align}
G(x,y) & = \sum_{n \geq 1} \sum_{k = 1}^n N(n,k) \,x^n  y^k \nonumber \\
& = \frac{1 + x(1-y) - \sqrt{1-2 x(1+y) + x^2 (1-y)^2} }{2x} - 1 . \label{eq:gen}
\end{align}
Then, letting
\begin{align}
x_\ast & = \pi(\up\vert \up) \pi(\down\vert \down) \\
y_\ast & = \frac{\pi(\up\vert \down) \pi(\down\vert \up)}{x_\ast},
\end{align}
we find that
\begin{align}
\mathfrak{f}_- = p^\mathcal{L}_1(\down) + p^\mathcal{L}_1(\up) \frac{\pi(\down\vert \down)}{\pi(\up\vert \down)}  G(x_\ast,y_\ast). \label{eq:fast}
\end{align}
Now notice that, using normalization of the trans-transition probabilities Eq.\,(\ref{eq:norm}), we have
\begin{align}
x_\ast (1 -y_\ast) & = \pi(\up \vert  \up) + \pi(\down \vert  \down) - 1 , \nonumber \\
x_\ast( 1+y_\ast) & = 2\pi(\up \vert  \up) \pi(\down\vert \down) - \pi(\up \vert  \up) - \pi(\down \vert  \down) + 1 . \label{eq:xy}
\end{align}
After some tedious but revealing calculation (see appendix \ref{app2} for details) one obtains that the square root in Eq.\,(\ref{eq:gen}) has real-valued solution
\begin{align}
\sqrt{1-2 x_\ast(1+y_\ast) + {x_\ast}^2 (1-y_\ast)^2} = \big \vert \pi(\up\vert \up) - \pi(\down\vert \down) \big\vert  \label{eq:sqrt}
\end{align}
in terms of the absolute value $\vert \,\cdot\,\vert $. We then find the remarkably simple expression
\begin{align}
G(x_\ast,y_\ast) = \left\{\begin{array}{ll}
\pi(\up\vert \down)/\pi(\down\vert \down), & \mathrm{if}\,\pi(\up\vert \up) < \pi(\down\vert \down) \\
\pi(\down\vert \up)/\pi(\up\vert \up), & \mathrm{if}\,\pi(\up\vert \up) \geq \pi(\down\vert \down) . 
\end{array} \right.  \label{eq:fss}
\end{align}
Plugging this latter into Eq.\,(\ref{eq:fast}) we find our central result
\begin{align}
\mathfrak{f}_- = \min\,\left\{ 1, p^\mathcal{L}_1(\down) + p^\mathcal{L}_1(\up) \frac{\pi(\down\vert \down)}{\pi(\up\vert \up)} \frac{\pi(\down \vert  \up)}{\pi(\up \vert  \down)} \right\}, \label{eq:inview}
\end{align}
where the two values are obtained respectively for $\pi(\up\vert \up) < \pi(\down\vert \down)$ and for $\pi(\up\vert \up) \geq \pi(
\down\vert \down)$. To express $\mathfrak{f}_-$ as a minimum between two values we used the fact that, because $\pi(\down \vert  \up)/ \pi(\up \vert  \down) = [1 - \pi(\up \vert  \up)]/[1- \pi(\down \vert  \down) ]$,  the second value is monotonically increasing in $\pi(\down\vert \down)$ and decreasing in $\pi(\up\vert \up)$, and is only $1$ for $\pi(\up\vert \up) = \pi(\down\vert \down)$.

Now notice that the stationary distribution in transition space (eigenvector of the trans-transition matrix relative to eigenvalue $1$, $Pp^\mathcal{L}_\infty = p^\mathcal{L}_\infty$) is easily found to be $p^\mathcal{L}_\infty(\ell) \propto \pi(\ell\vert \overline{\ell})$, where $\overline{\ell}$ denotes the reverse transition of $\ell$ (i.e. $\overline{\up} = \down$, $\overline{\down} = \up$). Therefore we can rewrite the above expression as
\begin{align}
\mathfrak{f}_- = \min\,\left\{ 1, p^\mathcal{L}_1(\down) + p^\mathcal{L}_1(\up) \frac{\pi(\down\vert \down) p^\mathcal{L}_\infty(\down)}{\pi(\up\vert \up) p^\mathcal{L}_\infty(\up)} \right\}. \label{eq:minus}
\end{align}

Now consider the probability $\mathfrak{f}_+$ that the cumulated  current ever reaches value $+1$. A quick review of the above derivation promptly leads to
\begin{align}
\mathfrak{f}_+ = \min\,\left\{p^\mathcal{L}_1(\up) + p_1^\mathcal{L}(\down) \frac{\pi(\up\vert \up) p^\mathcal{L}_\infty(\up)}{\pi(\down\vert \down) p^\mathcal{L}_\infty(\down)}, 1 \right\}, \label{eq:plus}
\end{align}
where the two values are taken respectively for  $\pi(\up\vert \up) < \pi(\down\vert \down)$ and for  $\pi(\up\vert \up) \geq \pi(\down\vert \down)$.

\subsection{The effective affinity}
\label{sec:effaff}


Let us define
\begin{align}
F & = \log \frac{\pi(\up\vert \up)}{\pi(\down\vert \down)}. \label{eq:ea}
\end{align}
This quantity has been given an operational thermodynamic interpretation in Refs.\,\cite{polettini2019effective,bisker2017hierarchical} as follows. By Eqs.\,(\ref{eq:explicit}) and (\ref{eq:ten}) we have
\begin{align}
F &= \log \frac{\rho(1\vert 2) [ \det R_{\setminus (2\vert 1)}  - 
\rho(2\vert 1) \det R_{\setminus (1,2 \vert  2, 1)}]}{
\rho(2\vert 1) [\det R_{\setminus (1\vert 2)}  -  
\rho(1\vert 2) \det R_{\setminus (1,2 \vert  2, 1)}]} \nonumber \\
& = \log \frac{\rho(1\vert 2) p_\infty^\varnothing(2) }{
\rho(2\vert 1)  p_\infty^\varnothing(1)} \label{eq:lolo}
\end{align}
where in the second expression $p^\varnothing_\infty = \lim_{t \to \infty} p^\mathcal{X_\varnothing}_{t}$ is the stationary probability of the system where transition $1 \leftrightarrow 2$ is removed, i.e. $R^\varnothing p^\varnothing_\infty = 0$ (see Appendix \ref{app2} for a direct proof; the distribution is unique by the assumption that edge $1 \leftrightarrow 2$ is not a bridge). Notice that this is a stalling system, that is, one where (by non-existence of the transition!) the mean stationary current $\langle \dot{c} \rangle^\varnothing = \rho(1\vert 2) p^\varnothing_\infty(2) - \rho(2\vert 1) p^\varnothing_\infty(1)$ vanishes.

Let us now parametrize rates according to the principle of local detailed balance \cite{maes2021local, esposito2010three}
\begin{align}
\frac{\rho(x\vert x')}{\rho(x'\vert x)} = \exp \frac{\delta q_{xx'}}{T_{xx'}}
\end{align}
in terms of a energy increment $\delta q_{xx'} = - \delta q_{x'x}$ and a temperature profile $T_{xx'} = T_{x'x}$ describing {\color{black} the influence of a local bath's degrees of freedom}. We assume that temperature $T_{12}$ is specific of transition $1 \leftrightarrow 2$ (that is, its variation does not affect other rates). Then it was proven \cite{polettini2019effective} that there exists a value $T^\varnothing_{12}$ for which the mean current stalls (but here the transition is possible!). Nevertheless, a simple argument shows that the stationary values $p^\varnothing_\infty(2)$ and $p^\varnothing_\infty(1)$ are the same as in the system where the transition is removed altogether (see Appendix \ref{app3}). We therefore have
\begin{align}
0 = \langle \dot{c} \rangle^\varnothing = \rho^\varnothing(1\vert 2) p^\varnothing_\infty(2) -  \rho^\varnothing(2\vert 1)  p^\varnothing_\infty(1)
\end{align}
leading to $p_\infty^\varnothing(2)/p_\infty^\varnothing(1) = -\delta q_{12} / T^\varnothing_{12}$ and
\begin{align}
F = \delta q_{12} \left( \frac{1}{T_{12}} - \frac{1}{T_{12}^\varnothing} \right).
\end{align}
This latter local expression grants {\color{black}an operational} procedure to measure $F$, on the assumption that $\delta q_{12}$ is {\color{black}measured or theoretically determined by a microphysical theory of the system describing energy levels}, that $T_{12}$ is tunable, and that the mean current $\langle \dot{c} \rangle$ is observable. {\color{black}The procedure consists in tuning $T_{12}$ to the value $T^\varnothing_{12}$ for which the observable mean current vanishes. Then, if $\delta q_{12}$ is known, $F$ is determined in terms of the inverse temperature difference.}

As regards the global acceptation of affinity mentioned in the introduction, for systems containing a single oriented cycle $\mathcal{C}$ it is easily shown \cite{harunari2022learn,van2022thermodynamic} that $F = A$ is the cycle affinity, namely the ratio of the  products of rates along the cycle, in opposite directions
\begin{align}
A = \log \prod_{(xx') \in \mathcal{C}} \frac{\rho(x\vert x')}{\rho(x'\vert x)} =  \sum_{(xx') \in \mathcal{C}} \frac{\delta q_{xx'}}{T_{xx'}} = \oint \frac{\delta q}{T}.
\end{align}
For vanishing $A$ (Kolmogorov condition) one finds an equilibrium state with vanishing mean current. From the above relation one immediately finds for the equilibrium temperature the relation
\begin{align}
\frac{\delta q_{12}}{T^\varnothing_{21}} = - \sum_{(12) \neq (xx') \in \mathcal{C}} \frac{\delta q_{xx'}}{T_{xx'}}.
\end{align}

For generic multicyclic systems, this latter identification with a specific thermodynamic cycle is not possible. However, the cumulated current $c = \sum_{\mathcal{C}} c({\mathcal{C})}$ can in fact be envisioned as the sum of the winding numbers over all cycles that include the visible transition (see Refs.\,\cite{polettini2021tight,jiang2022large} for some insights on such winding numbers). Notice that a stalling mean current does not imply global equilibrium, as these cycles may have circulation even if overall the visible mean current stalls. An explicit expression of $F$ in terms of such cycles is
\begin{align}
F = \log \frac{\sum_{\mathcal{C} \owns \up} w(\mathcal{C}) \prod_{(xx') \in \mathcal{C}} \rho(x\vert x')}{\sum_{\mathcal{C} \owns \up} w(\mathcal{C}) \prod_{(xx') \in \mathcal{C}} \rho(x'\vert x)}
\end{align}
where $w(\mathcal{C})$ is some cycle weight, independent of the cycle's orientation \cite{polettini2019effective}. Nevertheless, defining entropy production as the Kullback-Leibler distance of random processes from their time-reversed, it has been shown that $F \langle \dot{c} \rangle$ is indeed the entropy production estimated by an external observer who only has access to the sequence of visible transitions \cite{harunari2022learn,van2022thermodynamic}.

\subsection{Special cases and the noria, {\color{black} and a generalization}}

We consider two special cases where our main results write in terms of the effective affinity. Here we resolve the explicit dependency of the stopping probability in terms of the probability $p_1^\mathcal{L}$ of the first transition, $\mathfrak{f}_\pm = \mathfrak{f}_\pm\left[p_1^\mathcal{L}\right]$. Remember that such probability can eventually be computed from the initial probability in state space $p_0^\mathcal{X}$ via Eq.\,(\ref{eq:init}). {\color{black} Finally we generalize the above results to the probabilty of hitting arbitrary low values.}

\paragraph{Stationary case}

In the first case we sample the initial transition from the stationary distribution. We easily find from Eqs.\,(\ref{eq:minus}) and (\ref{eq:plus})
\begin{align}
\mathfrak{f}_-\left[p^\mathcal{L}_\infty\right] & =  \min\, \left\{1, p^\mathcal{L}_\infty(\down) (1 + \exp -F) \right\},\\
\mathfrak{f}_+\left[p^\mathcal{L}_\infty\right] & =  \min\, \left\{p^\mathcal{L}_\infty(\up) (1 + \exp +F), 1 \right\}.
\end{align}
From an operational point of view this is particularly simple because it only requires to wait long enough for the system to stationarize. Then  $p^\mathcal{L}_\infty$ can be computed explicitly from the time series of the transitions, by just counting the relative frequency of $\up$'s and $\down$'s.

\paragraph{Cyclic case}

In the second case, we prepare the system just after a visible transition is performed and then wait for the same transition to occur again, thus completing a cycle. Therefore for $c = +1$ we prepare the system at the tipping point of $\up$, which gives $p_0^\mathcal{X}(x) = \delta_{x,1}$ so that, after Eq.\,(\ref{eq:init}) is applied, $p^\mathcal{L}_1(\ell) = \pi(\ell\vert \up)$. For $c = -1$ we prepare the system at the tipping point of $\down$, which gives $p_0^\mathcal{X}(x) = \delta_{x,2}$ and $p^\mathcal{L}_1(\ell) = \pi(\ell\vert \down)$. After some calculation trick such as
\begin{align}
\pi(\up\vert \up) \left[ 1+ \frac{\pi(\up\vert \down)}{\pi(\down\vert \down)} \right] =  \pi(\up\vert \up) \left[ 1+  \frac{1- \pi(\down\vert \down)}{\pi(\down\vert \down)} \right] = \exp F
\end{align}
we find
\begin{align}
\mathfrak{f}_-\left[\pi(\cdot\vert \down)\right] & = \min\,\left\{1, \exp - F \right\}, \label{eq:ppm} \\
\mathfrak{f}_+\left[\pi(\cdot\vert \up)\right] & = \min\,\left\{ \exp + F , 1\right\}. \label{eq:ddm} 
\end{align}
This result is analogous to the one derived in Ref.\,\cite{bauer2014affinity} for unicyclic systems, with the exception that in the unicyclic case the choice of initial state (or, equivalently, the final transition) is not relevant, given that all states share the same cycle and therefore the explicit dependency on the initial state drops and the above result simplifies to
\begin{align}
\mathfrak{f}_\pm[\,\cdot\,] & = \min \,\left\{ 1, \exp \pm A \right\}
\end{align}
where $\mathfrak{f}_\pm[\,\cdot\,]$ is just the probability that the cycle is ever completed in either direction, independently of the initial state.

{\color{black} \paragraph{Hitting $-n$} The above hitting result for the cumulated current to ever become $-1$ (for $F > 0$) lends itself to a simple generalization to the case of the cumulated current hitting value $-n$, for $n \in \mathbb{N}$. Intuitively (given that denumerable + denumerable = denumerable) this is just given by reiterating the hitting problem (renewal property), with the initial condition stabilizing to the previous occurrence of $\down$ just after the first occurrence. One immediately obtains
\begin{align}
\mathfrak{f}_{-n}[p^{\mathcal{L}}_1] & = \mathfrak{f}_{-1}[p^{\mathcal{L}}_1] \; \mathfrak{f}_{-1}[\pi(\cdot\vert \down)]^{n-1}  = \left[ p^{\mathcal{L}}_1(\up) e^{F} + p^{\mathcal{L}}_1(\down) e^{ F^\leftrightarrow}  \right] e^{-nF}, \label{eq:fn}
\end{align}
where we rewrote $\pi(\up\vert \down)/\pi(\down\vert \up) = \exp F^{\leftrightarrow}$ as the effective affinity of a system whose trans-transition matrix $P^\leftrightarrow$ has the columns swapped with respect to $P$; interestingly this auxiliary dynamics also plays a role in formulating the transient fluctuation relation in Ref.\,\cite{harunari2022beat}, but its physical interpretation has still to be clarified.}

\subsection{Fluctuation relations}

In the unicyclic case, one easily finds the fluctuation relation
\begin{align}
\frac{\mathfrak{f}_+[\,\cdot\,] }{\mathfrak{f}_-[\,\cdot\,]} = \exp A.
\end{align}

In the multicylic case, from Eqs.\,(\ref{eq:ddm}, \ref{eq:ppm}) we have
\begin{align}
\frac{\mathfrak{f}_+\left[\pi(\cdot\vert \up)\right] }{\mathfrak{f}_-\left[\pi(\cdot\vert \down)\right] } = \exp F . \label{eq:estimator}
\end{align}
This looks formally like a fluctuation relation, with a {\it caveat}: in fluctuation relations the probabilities being compared should be the same, while in this case they are different probabilities, as they are conditioned on two different initial distributions, viz. $\pi(\cdot\vert \up)$ and $\pi(\cdot\vert \down)$. This, as we will see, has consequences on the computational or experimental interpretation of data, given that one should prepare different experiments for forward and backward processes and post-select their outcome, which is not desirable. In the next section we comment further on this aspect, arguing that Eq.\,(\ref{eq:estimator}) may in fact be the best chance of an estimator of nonequilibrium despite approximations.

Furthermore, in Ref.\,\cite{harunari2022beat} it was proven (Eq.\,(21)) that, by sampling the initial transition from distribution $p_1^\mathcal{L}(\ell) \propto \pi(\ell\vert \ell)$, the following fluctuation relation holds
\begin{align}
\frac{p_n(c)}{p_n(-c)} = \exp c F, 
\end{align}
{\color{black} where we remind that $c$ is given by Eq.\,(\ref{eq:cur}) and} $p_n(c)$ is the probability that the cumulated current is a certain value $c \in \mathbb{Z}$ after $n$ visible transitions. One can then further derive the relation
\begin{align}
\frac{\sum_{n \in \mathcal{N}} p_n(+1)}{\sum_{n  \in \mathcal{N}} p_n(-1)} = \exp F  \label{eq:fr2}
\end{align}
where $\mathcal{N}$ is any subset of $\mathbb{N}$. This is reminiscent of Eq.\,(\ref{eq:estimator}), but notice that these latter are not independent probabilities.

Finally, fluctuation relations for single edge currents at stopping times different than the total number of visible transitions (in particular at ``clock time'' $t$) do not generally hold -- but in the unicyclic case -- because the statistics of a specific current depends on all other currents flowing through the network. This is what makes relations such as Eqs.\,(\ref{eq:estimator}) and (\ref{eq:fr2}) particularly appealing, as they are local and phenomenological, and do not depend on knowledge of the whole system.

\subsection{Estimation of the effective affinity}

Many of the above expressions can be used to build estimators of the effective affinity. We will focus on the ones coming from cyclic processes.

Consider $M$ independent realizations of a trajectory performing $N$ visible transitions:
\begin{align}
\ell_1^{(m)} ,\ell_2^{(m)},\ldots,\ell_{N}^{(m)}, \quad m \in [1,M] .
\end{align}
Define the cumulated current after the $n$-th visible transition
\begin{align}
\hat{c}^{(m)}_n = \sum_{k = 1}^n \left( \delta_{\ell_k^{(m)}, \up} - \delta_{\ell_k^{(m)}, \down}\right).
\end{align}
It has empirical distribution
\begin{align}
\hat{p}_n(c) = \sum_{m = 1}^M \delta_{\hat{c}^{(m)}_n, c}, \quad \mathrm{for} \, c \in [-n,n]
\end{align}
and empirical mean and variance
\begin{align}
\langle \hat{c}_n \rangle & = \frac{1}{M} \sum_{m = 1}^M  c^{(m)}_n = \sum_{c \in [-n,n]} c \, \hat{p}_n(c)  , \\
\langle\!\langle \hat{c}_n^2 \rangle\!\rangle & = \frac{1}{M} \sum_{m = 1}^M  (\hat{c}^{(m)}_n - \langle \hat{c}_n \rangle  )^2 = \sum_{c \in [-n,n]} c^2 \, \hat{p}_n(c) - \langle \hat{c}_n \rangle^2.
\end{align}
Define the empirical stopping times
\begin{align}
\hat{N}_{\pm}^{(m)} = \inf  \{n \in [0,N] \;\mathrm{s.t.}\; \hat{c}^{(m)}_n = \pm 1\} \lor \{N+1\}
\end{align}
and the estimators of the stopping probabilities
\begin{align}
\hat{\mathfrak{f}}_\pm & = \min \left\{ 1 - \frac{1}{M} \sum_{m = 1}^M \delta_{\hat{N}_{\pm}^{(m)}, N+1}, \frac{1}{M} \right\}
\end{align}
where the minimum is introduced to avoid possible divergences in the case $\hat{\mathfrak{f}}_\pm = 0$ (see also Eq.\,(25) in Ref.\,\cite{neri2022estimating}).

Notice that due to the finite cutoff on the number of transitions, given Eq.\,(\ref{eq:sum}) these latter are biased. In particular they systematically underestimate (on average) the true stopping probability due to the fact that all occurrences of $c = \pm 1$ after $N$ visible transitions are discarded. 

\begin{figure}[t]
\begin{center}
\includegraphics[width=10cm]{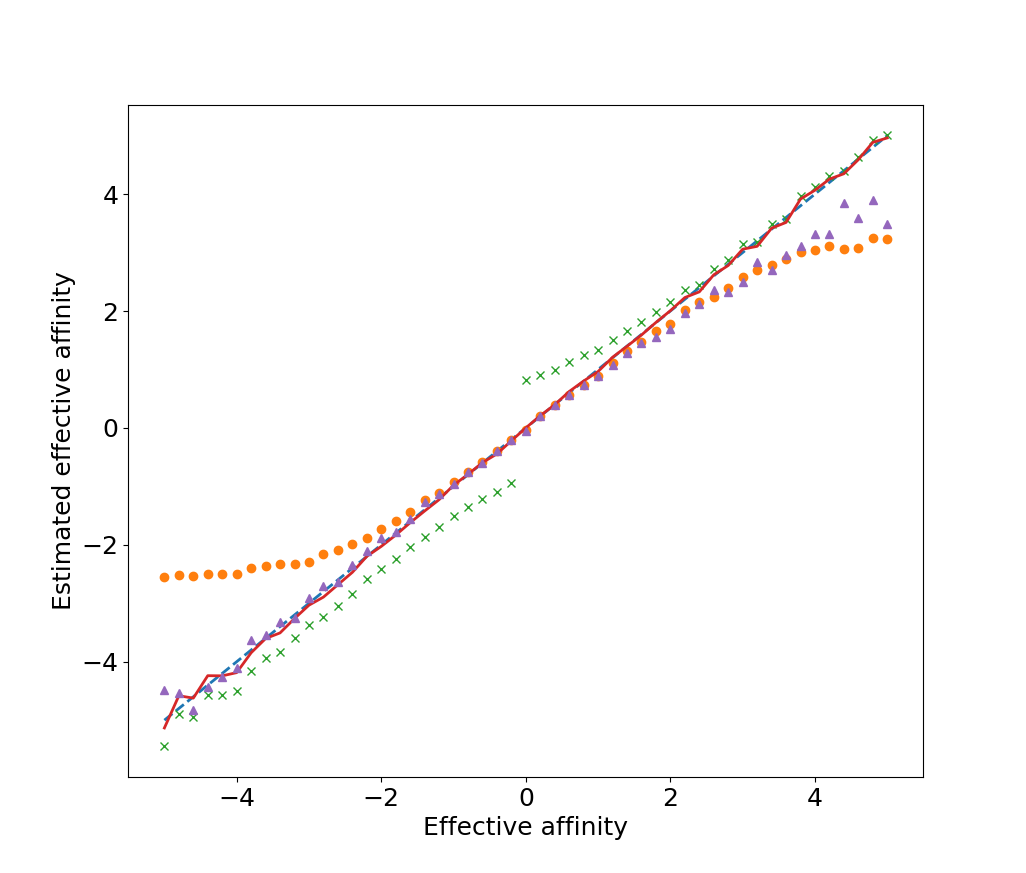}
\caption{For a fully-connected four-state model with all unit rates except for $R_{1,2} = \exp F$ and initial state $x=1$ (that is $p^{\mathcal{X}}_0(x) = \delta_{1,x}$): {\it (dashed)} the effective affinity $F$; {\it (continuous)} estimator $\hat{F}_{\mathrm{fr}}$ of $\log \mathfrak{f}_+/ \mathfrak{f}_-$; {\it (crossed)} estimator $\hat{F}_{\mathrm{cy}}$ of $\mathrm{sign}\, (\mathfrak{f}_+ - \mathfrak{f}_-) \log \mathfrak{f}_\sigma$; {\it (bullets)} the linear regime estimator $\hat{F}_{\mathrm{lr}}$; {\it (triangles)} the entropy production estimator $\hat{F}_{\mathrm{kl}}$. The ultimate stopping time was set to $N = 20$ and the number of samples to $M = 10000$.}
\end{center}
\end{figure}

Assuming that we can ignore the initial conditions, we can invert Eqs.\,(\ref{eq:ppm}) and (\ref{eq:ddm}) to obtain an estimator of the effective affinity
\begin{align}
\hat{F}_{\mathrm{cy}} & = \left\{ \ba{ll}
\log \hat{\mathfrak{f}}_+ , & \mathrm{if\;} \hat{\mathfrak{f}}_+ \leq \hat{\mathfrak{f}}_-, \\ 
-\log \hat{\mathfrak{f}}_- , &  \mathrm{if\;} \hat{\mathfrak{f}}_+ > \hat{\mathfrak{f}}_- .
\ea \right.
\end{align}
We can compare this to the estimator coming from the stopping fluctuation relation
\begin{align}
\hat{F}_{\mathrm{fr}} & = \log \hat{\mathfrak{f}}_+ - \log \hat{\mathfrak{f}}_-,
\end{align}
which is generally biased due to the different initial conditions in Eqs.\,(\ref{eq:estimator}).

We complement these stopping-problem estimators with an estimator coming from the theory of linear response out of stalling states \cite{altaner2016fluctuation}
\begin{align}
\hat{F}_{\mathrm{lr}} & = \frac{2 \langle \hat{c}_N \rangle}{\langle\!\langle \hat{c}_N^2 \rangle\!\rangle }
\end{align}
and with an estimator obtained from the standard entropy production expression as a Kullback-Leibler divergence (properly regularized to avoid taking $\log 0$)
\begin{align}
\hat{F}_{\mathrm{kl}} & = \frac{1}{ \langle \hat{c}_N \rangle}\sum_{\substack{c \in [-N,N] \\ \hat{p}_N(c)\hat{p}_N(-c) \neq 0}} \hat{p}_N(c) \log  \frac{\hat{p_N}(c)}{\hat{p}_N(-c)}.
\end{align}
This latter is well-known to be a biased estimator, and better practices in evaluating relative entropies correct these biases but also greatly increase the running time (see Supplementary Material in Ref.\,\cite{harunari2022learn}). We do not concern ourselves with this issue here.

In Fig.\,2 we compare the behaviour of these estimators in a simple model. The linear regime estimator $\hat{F}_{\mathrm{lr}}$ performs better near the stalling condition $F = 0$, while it diverges significantly out of stalling. On the contrary, the cyclic estimator $\hat{F}_{\mathrm{cy}}$ converges far from stalling, but it systematically suffers from the finite $N$ cutoff. The entropy production estimator $\hat{F}_{\mathrm{kl}}$ is also  biased and noisy due to the tails of the cumulated current's distribution. The stopping fluctuation-relation estimator $\hat{F}_{\mathrm{fr}}$ instead appears to not be affected by all these issues, despite the approximation due the bias due to the different initial state in Eq.\,(\ref{eq:estimator}).

{\color{black} The reason behind the left-right asymmetry in the above plot is due to the fact that for simplicity we decided to only perturb one rate $R_{1,2} = \exp F$ and keep all others fixed. This choice is useful to show that entropy production estimators can lead to noise in the tails depending on time-scale separation between rates. Had we distributed the perturbation among $R_{1,2}$ and $R_{2,1}$, we would have obtained a more symmetric plot.}

\section{Discussion}

\subsection{{\color{black} A nonequilibrium Boltzmann formula?}}

{\color{black} Our Eq.\,(\ref{eq:norianoria}) can be seen as a ``nonequilibrium Boltzmann formula'' given its similarities with  $S = \log W$ connecting entropy $S$ and probability $1/W$ ($W$ being the volume of state space), elaborated by Boltzmann and refined by Planck (Boltzmann's constant set to unity). But with some precautions.}

Einstein wrote about the Boltzmann formula: \guillemotleft To be able to calculate $W$, one needs a complete theory of the system under consideration. If considered from a phenomenological point of view [this] equation appears devoid of content\guillemotright. Einstein then inverted the equation to make it a rule for inferring probabilities from measured entropy differences between equilibrium states -- which better capture the dynamical nature of processes -- and used it to perfection Smoluchowski's theory of critical opalescence \cite{jona2015large}. However, at least since Kant, philosophers warn us that observations are not independent of conceptions, and therefore deduction from measurements needs theory (the fluctuations of {\it what}?), and theory needs the human touch. Still now we don't know which came first, whether the chickens of gases and thermal machines or the eggs of thermodynamics and statistical mechanics \cite{penocchio}. At equilibrium the situation is aggravated by the fact that the construction of thermodynamic potentials requires many arbitrary choices by the observer \cite{polettini2013dice}, while the pursue of objectiveness requires a description of processes in terms of invariant quantities.

Far from equilibrium, flows of heat to and from the environment are not quantified by differences of a state function, but by ``inexact differences''. By the so-called principle of local detailed balance ratios of probabilities of forward-to-backward processes have been connected to so-called affinities that quantify the entropy production along cyclic processes, and which are invariant upon the redefinition of the fundamental degrees of freedom \cite{polettini2013dice}. However, until recently it has proven difficult to directly connect probabilities and meaningful physical quantities. In fact, despite some claims, there are no predictive variational principles far from equilibrium\cite{maes2007minimum,polettini2013fact}.

{\color{black}This latter issue is solved by our equations, which allow to connect a statistical property and a physical object in a more direct way than did, for example, fluctuation relations. However, the former concern still plagues our result. What comes first: the egg of $F$, or the chicken of $\mathfrak{f}_-$? Only circumstances can tell.}

\subsection{Relation to a companion publication}

The present manuscript is strictly related to a companion work \cite{neri2022extreme} by the same Authors that addresses similar questions. Let us clarify in which ways.

Equation (\ref{eq:fn}) is strictly related to Eq.\,(14) in Ref.\,\cite{neri2022extreme}. There the normalized probability $\mathfrak{p}_{-n}$ of the cumulated current taking minimum value $-n$ is addressed, while in our case $\mathfrak{f}_{-n}$ allows that, after hitting value $-n$, the cumulate current may take even more negative values. Therefore we have, intuitively, that this latter is the cumulative distribution of the former
\begin{align}
\mathfrak{f}_{-n} = \sum_{k \geq n} \mathfrak{p}_{-k}.
\end{align}
Given that $\mathfrak{p}_{-k}$ is normalized, this identification allows to estimate the escape probability that the cumulated current never actually attains a negative value as $\mathfrak{p}_0 = 1 - \mathfrak{f}_{-1}$, which in view of Eqs.\,(\ref{eq:inview}), the explicit expression for the trans-transition probabilibites Eqs.\,(\ref{eq:explicit}) and (\ref{eq:ten}), and the explicit expression for the probability of the first transition Eq.\,(\ref{eq:init}) allows to express $\mathfrak{p}_0$ in terms of the (distribution of) the initial state (see below the explicit expression).

The other main difference between the two works is methodological. Here we follow a constructive but specific approach based on first-transition time techniques and combinatorics, while Ref.\,\cite{neri2022extreme} is rooted in the more general theory of martingales. In particular in  Ref.\,\cite{neri2022extreme} it is shown that, upon a proper choice of initial state, $\exp -Fc$ is a martingale, and in particular its expected value $\langle \exp -Fc \rangle$ is constant in time. Doob's optional stopping theorem then states that this time can be any proper stopping time. By choosing the moment when the cumulated current hits the boundary values $n_+ > 0$ or $n_- < 0$ for the first time, and given that $c$ starts from value $0$, one obtains
\begin{align}
1 = \left\langle \exp -Fc \right\rangle = \mathfrak{f}^{(n_-)}_{n_+} e^{-Fn_+} + \mathfrak{f}^{(n_+)}_{n_-} e^{-Fn_-}
\end{align}
where $\mathfrak{f}^{(n_+)}_{n_-}$ is the probability of hitting $n_-$ whilst not hitting $n_+$. The nonequilibrium Boltzmann formula follows by taking $n_- = -1$ and $n_+ \to \infty$, in which limit $\mathfrak{f}^{(n_+)}_{-1} \to \mathfrak{f}_{-1}$. Interestingly, similar formulas were derived in an optimization context in Ref.\,\cite{cavina2016optimal}.

Finally, here is a short dictionary of equivalent terms and concepts in the two papers: transition rates $\rho(x\vert x')$ here are $k_{u \to v}$ there; the observed edge $1 \leftrightarrow 2$ is $y \leftrightarrow x$; ``cumulated'' currents $c$ are ``integrated'' currents $J$; for stationary probabilities we have $\infty$ instead of ``$\mathrm{ss}$''; the effective affinity $F$ is $a^\ast$; the extremum probabability $\mathfrak{p}_{-n}[p^\mathcal{L}_1]$, given Eq.\,(\ref{eq:init}), is $p_{J^{\mathrm{inf}}_{x \to y}} (-\ell \vert  X(0) = x_0)$; the escape probability $\mathfrak{p}_0$ is 
\begin{align}
p_{\mathrm{esc}}(x_0) = 1 + \left(  k_{x \to y}  \frac{   [S^{-1}]_{x,y}  [S^{-1}]_{y,x_0}}{   [S^{-1}]_{y,x}  }   +   k_{y \to x}  \frac{   [S^{-1}]_{y,y}  [S^{-1}]_{x,x_0}}{   [S^{-1}]_{x,x}  } \right)  e^{-a^\ast} \label{eq:pesc}
\end{align}
where $S$ is the matrix with entries $S_{u,v} = k_{v \to u} - \delta_{u,v} \sum_{w \in \mathcal{X}; w \neq u} k_{u \to w}$ if $(u,v) \neq (x,y), (y,x)$, else $S_{x,y} = S_{y,x} = 0$, and we used the explicit expression of the effective affinity Eq.\,(\ref{eq:ea}), that now translates into
\begin{align}
a^\ast = \log \frac{k_{x \to y} [S^{-1}]_{x,y}}{k_{y \to x} [S^{-1}]_{y,x}}.
\end{align}
When $x_0 = x$ we find
\begin{align}
p_{\mathrm{esc}}(x) & = 1 + \left( k_{x \to y}[S^{-1}]_{x,y} +   k_{y \to x}  [S^{-1}]_{y,y} \right) e^{-a^\ast} \nonumber \\
& = 1 - \left( \frac{ k_{x \to y} \det S_{\setminus (y\vert x)}  - k_{y \to x} \det S_{\setminus (y\vert y)}}{\det S}  \right) e^{-a^\ast} \nonumber \\
& = 1 - e^{-a^\ast},
\end{align}
where this latter passage follows from the algebraic manipulations in Appendix \ref{app1}. We thus recover Eq.\,(16) in the companion paper. We checked computationally the more general equivalence (implied by the theory) of Eq.\,(\ref{eq:pesc}) with Eq.\,(80) in the companion paper, but a direct proof  has remained elusive.

\subsection{Conclusions} 

Both martingale and first-transition methods are having a revival in connection to thermodynamic considerations \cite{harunari2022beat,harunari2022learn,sekimoto2021derivation,van2022thermodynamic,neri2017statistics,neri2019integral,neri2022universal}, and they may lead to independent generalizations and applications of our results. In both approaches, the main open question is the generalization to an arbitrary subset of currents -- neither the full entropy production nor a single edge current.

As regards the first-transition approach followed here, as soon as one steps out of the single-edge case the Markov property of the process in transition space is lost. Here the combinatorial approach may allow some exploration.

{\color{black} Since any Radon-Nicodym derivative of two probability distributions over realizations of the process is a martingale, martingales can be used to generalise the results in this paper. From this approach it would seem that one can generate an arbirary number of first-hitting results by building {\it ad hoc} auxiliary dynamics. However, the physical interpretation of this class of results may not be clear: it is crucial in our approach that the effective affinity has a clear operational interpretation. In particular, if one could tune its value by just ``turning a knob'', then the effective affinity is just the difference of that knob's value  (in proper physical units) where one wants to perform the experiment and the value of that knob at which the observable current vanishes on average. This local operational interpretation dispenses one to compute the effective affinity from knowledge of all the inner details of the fundamental thermodynamic cycles that influence that particular current.}

On a more speculative side, notice that in our derivation we made an arbitrary restriction of the solution of the {\color{black} Narayana} generating function, based on the assumption that we expect probabilities to be real-valued. It may be interesting to explore the meaning of the complex-valued solution.

Notoriously, Boltzmann's epitaph is his formula. But it took a whole community (including Einstein, Planck etc.) to digest it. So who's formula is it?

\section*{Data availability statement}
Data sharing not applicable to this article as no datasets were generated or analysed during the current study.

\section*{Conflict of interest statement}
The research was supported by the National Research Fund Luxembourg (project CORE ThermoComp C17/MS/11696700) and by the European Research Council, project NanoThermo (ERC-2015-CoG Agreement No. 681456).

\section*{Acknowledgements} MP is grateful to Paulo Fernando L\'evano for a philosophical consultation.

\bibliography{bibliography}

\appendix

\section{Appendices}

\subsection{Trans-transition probabilities in terms of minors}
\label{app1}

We prove Eqs.\,(\ref{eq:explicit}), given (\ref{eq:trantrans}) and (\ref{eq:ten}). We use the well-known matrix inverse
\begin{align}
[S^{-1}]_{x,x'} = (-1)^{x+x'}  \frac{\det S_{\setminus (x'\vert  x)} }{\det S},
\end{align}
where we remind that $S$ (as per Eq.\,(\ref{eq:S})) is a matrix obtained from the generator $R$ (Eq.\,(\ref{eq:generator})) by setting to zero the off-diagonal entries $(1,2)$ and $(2,1)$.

First consider $\det S_{\setminus (1\vert  1)}$ and $\det S_{\setminus (2\vert  2)}$. Since the removal of the first line and column, and of the second line and column, both take away the entries $(1,2)$ and $(2,1)$ which are the only ones that differ among $S$ and $R$, these determinants are identical to $\det R_{\setminus (1\vert  1)}$ and $\det R_{\setminus (2\vert  2)}$. Therefore from Eq.\,(\ref{eq:trantrans}) we find
\begin{align}
\pi(\up\vert  \down) & = - R_{1,2} [S^{-1}]_{2, 2} \nonumber \\
& = - \frac{R_{1,2} \det R_{\setminus (2\vert  2)}}{\det S} \nonumber \\
& = \frac{R_{1,2} \det R_{\setminus (2\vert  1)}}{\det S} \nonumber \\
& = \frac{\nu_\up}{\det S} \label{eq:similar}
\end{align}
where in the second passage we used the well-known fact that for any stochastic rate matrix $R$ the cofactors $(-1)^{x+x'} \det R_{\setminus (x\vert  x')}$ are independent of $x'$ (this follows for example from Eq.\,(\ref{eq:lala}) in appendix \ref{app3}), and in the third we used the definition in Eqs.\,(\ref{eq:explicit}) and (\ref{eq:ten}). A similar formula is found for $\pi(\up\vert  \down)$.

As regards $\det S_{\setminus (1\vert  2)}$ (respectively, $\det S_{\setminus (2\vert  1)}$), in this case the matrix resulting from the removal of the first row and second column only differs from $R_{\setminus (1\vert  2)}$ by entry $R_{2,1}$. Using the Laplace cofactor expansion for determinants we thus obtain
\begin{align}
\det S_{\setminus (1\vert  2)} = \det R_{\setminus (1\vert  2)} - 
\rho(1\vert  2) \det R_{\setminus (1,2 \vert   2,1)} \label{eq:dede}
\end{align}
and given the definitions in Eq.\,(\ref{eq:ten}) similar formulas as Eq.\,(\ref{eq:similar}) follow for $\pi(\up\vert  \up)$ and $\pi(\down\vert  \down)$.

Finally, consider any matrix $A$, and decompose its determinant in terms that are respectively linear in $A_{1,2}$, linear in $A_{2,1}$ or that contain $A_{1,2} A_{2,1}$
\begin{align}
\det A = a  A_{1,2} + b A_{2,1} + c A_{1,2} A_{2,1} + d \label{eq:detdet}
\end{align}
where $a, b,c,d$ are four parameters that do not depend on $A_{1,2}$ and $A_{2,1}$ and are unrelated to previous notation. By Jacobi's formula the dependency on $A_{1,2}$ is
\begin{align}
\frac{\partial \det A}{\partial A_{1,2}} = - \det A_{\setminus (1\vert  2)} . \label{eq:par}
\end{align}
Repeating with respect to $A_{2,1}$ we obtain
\begin{align}
c = \frac{\partial^2 \det A}{\partial A_{1,2} \partial A_{2,1}} = \det A_{\setminus (1,2\vert  1,2)}.
\end{align}
Coefficient $a$ is found from Eq.\,(\ref{eq:par}) by subtracting away this latter term multiplied by $A_{2,1}$, and similarly for $b$:
\begin{align}
a & = - \det A_{\setminus (1\vert  2)}  - A_{2,1} \det A_{\setminus (1,2\vert  1,2)} \\
b & = - \det A_{\setminus (2\vert  1)}  - A_{1,2} \det A_{\setminus (1,2\vert  1,2)}. 
\end{align}
Taking $A = R$, we have $\det A = 0$ and $d = \det S$. Therefore we obtain
\begin{align}
\det S & = d - \det A  \nonumber \\
& = - a A_{1,2} - b A_{2,1} - c A_{1,2} A_{2,1} \nonumber  \\
& = 
\rho(1\vert  2) \det R_{\setminus (1\vert  2)} + 
\rho(2\vert  1) \det R_{\setminus (2\vert  1)}+ 
\rho(1\vert  2) 
\rho(2\vert  1)  \det R_{\setminus (1,2\vert  1,2)}
\end{align}
where in the first passage we used Eq.\,(\ref{eq:detdet}) and in the second we used the explicit expressions for the paramters. In view of Eqs.\,(\ref{eq:ten}) this yields the trans-transition probabilities in Eq.\,(\ref{eq:explicit}). $\Box$

\subsection{From the {\color{black} Narayana} generating function to trans-transition probabilities}
\label{app2}

First let us show Eq.\,(\ref{eq:sqrt}). Using Eqs.\,(\ref{eq:xy}) we have
\begin{align}
& 1-2 x_\ast(1+y_\ast) + {x_\ast}^2 (1-y_\ast)^2 \nonumber \\
& = 1 - 2 [ 2\pi(\up \vert   \up)\pi(\down\vert  \down) - \pi(\up \vert   \up) - \pi(\down \vert   \down) + 1 ] +  [\pi(\up \vert   \up) + \pi(\down \vert   \down) - 1]^2  \nonumber \\
& = 1 - 4\pi(\up \vert   \up)\pi(\down\vert  \down) + 2\pi(\up \vert   \up) + 2\pi(\down \vert   \down) - 2  \nonumber \\
& \phantom{=} + \pi(\up \vert   \up)^2 + \pi(\down \vert   \down)^2 + 1 + 2 \pi(\up\vert  \up) \pi(\down\vert  \down) - 2 \pi(\up\vert  \up) - 2 \pi(\down \vert  \down)  \nonumber \\
& = [\pi(\up \vert   \up) - \pi(\down \vert   \down) ]^2.
\end{align}
Then from Eq.\,(\ref{eq:gen}), for $\pi(\up\vert  \up) \geq \pi(\down\vert  \down)$
\begin{align}
G(x_\ast,y_\ast) & = \frac{\pi(\up \vert   \up) + \pi(\down \vert   \down) - \pi(\up \vert   \up) + \pi(\down \vert   \down) }{2 \pi(\up\vert  \up) \pi(\down\vert  \down)} - 1 \\
& = \frac{1}{\pi(\up\vert  \up)} - 1 \\
& = \frac{\pi(\down\vert  \up)}{\pi(\up\vert  \up)},
\end{align}
which is the lower entry of Eq.\,(\ref{eq:fss}). Similarly for the other case.

\subsection{Effective affinity and stalling distribution}
\label{app3}

First notice that the stationary distribution $p^\mathcal{X}_\infty(x)$ of a continuous-time Markov generator $R$ can be found as follows. Given that $\det R = 0$, expanding with Laplace's cofactor formula, we find
\begin{align}
0 = \det R = \sum_{x} (-1)^{x'+x} R_{x'x} \det R_{\setminus (x'\vert  x)}. \label{eq:lala}
\end{align}
But then
\begin{align}
p^\mathcal{X}_\infty(x) = Z^{-1} (-1)^{x+x'} \det R_{\setminus (x'\vert  x)}
\end{align}
for any choice of $x'$, where $Z$ is the normalization.

Now let $R^\varnothing$ be the generator of the continuous-time Markov process where the visible rates are set to zero, $
\rho(1\vert  2) = \rho(2\vert  1) = 0$. We obtain, choosing $(x,x') = (1,2)$ and $(x,x') = (2,1)$
\begin{equation}
\begin{aligned}
p^\varnothing_\infty(1) = - Z^{-1} \det R^\varnothing_{\setminus (2\vert  1)} \\
p^\varnothing_\infty(2) = - Z^{-1} \det R^\varnothing_{\setminus (2\vert  1)}
\end{aligned}
\end{equation}
where the second follows from Eq.\,(\ref{eq:dede}). But now notice that removing the first row and secondo column [or viceversa] from $R^\varnothing$ results in the same matrix as by removing the first row and second column [or viceversa] from $S$. Therefore in view of Eq.\,(\ref{eq:dede}), and because of the cancellation of the terms $- Z^{-1}$, we find Eq.\,(\ref{eq:lolo}). $\Box$

Finally, consider a system with local rates tuned to a stalling temperature $T^\varnothing$ according to the principle of local detailed balance
\begin{align}
\frac{\rho^{\varnothing}(1\vert  2)}{\rho^{\varnothing}(2\vert  1)} = \exp \frac{\delta q_{12}}{T^\varnothing}
\end{align}
such that the mean current vanishes. Let $R^{T_\varnothing}$ be its generator. Notice that it differs from $R^\varnothing$. However, its stationary distribution is the same. In fact, computing $R^{T_\varnothing}_{x,x'} p_\infty^\varnothing(x')$ we find
explicitly
\begin{align}
\forall x \neq 1,2 , \qquad \sum_{x'}\left[ \rho(x\vert  x') p_\infty^\varnothing(x') - \rho(x'\vert  x) p_\infty^\varnothing(x) \right] & = \sum_{x'} R^\varnothing_{x,x'} p_\infty^\varnothing(x')  \nonumber \\
\sum_{x' \neq 2} \left[ 
\rho(1\vert  x') p_\infty^\varnothing(x') - \rho(x'\vert  1) p_\infty^\varnothing(1) \right] + \stackrel{= 0}{\overbrace{\rho^\varnothing(1\vert  2) p_\infty^\varnothing(2) - \rho^\varnothing(2\vert  1) p_\infty^\varnothing(1)}}  & 
= \sum_{x'} R^\varnothing_{x,x'} p_\infty^\varnothing(x')  \nonumber \\
\sum_{x' \neq 1} \left[ 
\rho(2\vert  x') p_\infty^\varnothing(x') - \rho(x'\vert  2) p_\infty^\varnothing(2) \right] + \stackrel{= 0}{\overbrace{\rho^\varnothing(2\vert  1) p_\infty^\varnothing(1) - \rho^\varnothing(1\vert  2) p_\infty^\varnothing(2)}}  & 
= \sum_{x'} R^\varnothing_{2,x'} p_\infty^\varnothing(x') 
 \end{align}
where we used the fact that the mean current vanishes by assumption, and on the right-hand side we recognized the stationary equation $R^\varnothing p_\infty^\varnothing = 0$. $\Box$.

\end{document}